\documentclass[sigconf]{acmart}

\usepackage{booktabs}
\usepackage{multirow}
\usepackage{graphicx}

\AtBeginDocument{%
  }

\setcopyright{acmlicensed}
\copyrightyear{2026}
\acmYear{2026}
\acmDOI{XXXXXXX.XXXXXXX}
\acmConference[CIKM '26]{The 35th ACM International Conference on Information and Knowledge Management}{November 7--11, 2026}{Rome, Italy}
\acmISBN{978-1-4503-XXXX-X/2026/05}

\begin{document}




\title{Graph-GRPO: Dependency-Aware Credit Assignment for Generative E-commerce Search Relevance}
\settopmatter{authorsperrow=3}

\author{Jiarui Che}
\authornote{Both authors contributed equally to this research.}
\email{2120240716@mail.nankai.edu.cn}
\orcid{1234-5678-9012}
\affiliation{%
  \institution{Nankai University}
  \city{Tianjin}
  \country{China}
}

\author{Yifei Chen}
\authornotemark[1]
\email{chenyifei44@jd.com}

\author{Zhixing Tian}
\email{tianzhixing1@jd.com}
\affiliation{%
  \institution{JD.COM}
  \city{Beijing}
  \country{China}
}

\author{Chenyang Wang}
\email{wangchenyang3@jd.com}

\author{Ziguang Cheng}
\email{chengziguang1@jd.com}
\affiliation{%
  \institution{JD.COM}
  \city{Beijing}
  \country{China}
}






\renewcommand{\shortauthors}{Che et al.}

\begin{abstract}
Search relevance modeling is a core task in e-commerce search systems, assessing how well a user query matches candidate products. Rather than relying on a single holistic matching signal, relevance judgment often requires structured reasoning over query understanding, product understanding, and facet-level matching. With large language models (LLMs), this process is increasingly formulated as chain-of-thought (CoT) reasoning and optimized with reinforcement learning (RL). However, existing RL methods mainly rely on outcome-level rewards and treat the entire reasoning chain as a single optimization unit. This makes it difficult to distinguish faulty reasoning steps from correct intermediate ones, leading to misaligned credit assignment. Although process-reward methods provide denser supervision, they often treat reasoning steps independently and ignore dependency-driven error propagation, making responsibility attribution difficult and limiting the optimization of structured relevance reasoning.

We propose Graph-GRPO, a graph-structured extension of GRPO for multi-component relevance reasoning. Graph-GRPO constructs a relevance reasoning dependency graph, where CoT steps are modeled as nodes and their logical dependencies as edges. It propagates outcome-level rewards over the graph to derive step-level credit signals, enabling more accurate fine-grained credit assignment. We further introduce a main-loss-driven controller that adaptively adjusts edge-wise credit-propagation coefficients. Together with CoT random masking for supervised policy initialization and graph-node-based multi-head distillation, we build a trainable and deployable framework for generative relevance modeling. Extensive offline evaluations and online A/B tests on a leading e-commerce platform demonstrate that the Graph-GRPO-based framework improves relevance classification metrics and key engagement metrics. The resulting lightweight model has been deployed in the JD e-commerce search system, serving hundreds of millions of users.
\end{abstract}




\maketitle

\section{Introduction}

E-commerce search serves as a critical entry point that connects user shopping needs with product supply, where the key challenge lies in accurately assessing the relevance between a user query and a candidate product. As a fundamental task in e-commerce search systems, relevance modeling directly affects user experience and platform conversion efficiency. Traditional relevance models are mostly based on BM25 \cite{robertson1994okapi}, vector matching \cite{huang2013learning,yi2019sampling}, or BERT-style discriminative architectures \cite{devlin2019bert,liu2019roberta,lan2019albert}, directly predicting a relevance label or score for each query-product pair. However, in practice, relevance judgment is often not a simple text-matching problem. It requires parsing the subject and attribute information in both the query and the product, and comprehensively assessing their matching relationship. Generative relevance modeling based on large language models provides stronger semantic understanding and reasoning capabilities \cite{pradeep2023rankvicuna,zhuang2023rankt5,sun2023chatgpt,sachan2022improving}, which helps model complex matching relations and pushes relevance modeling from single-step label prediction toward a structured reasoning paradigm \cite{tang2025lref,dong2025taosr1,yifei2026k,fang2025adore}.

As relevance modeling shifts toward structured relevance reasoning, optimizing the reasoning process itself becomes increasingly important. RL provides a natural optimization paradigm for this setting, as it can refine the generation policy with task-specific feedback \cite{schulman2017proximal,shao2024deepseekmath}. However, existing methods \cite{tang2025lref,dong2025taosr1,zeng2026optimizing} usually rely on sparse outcome-level rewards and treat the entire reasoning chain as a single optimization unit. This makes it difficult to accurately attribute responsibility to different reasoning steps. Although process rewards constructed from manual rules or external reward models can provide denser supervision \cite{xia2026reasoning}, they are costly to construct and often unstable in practice. More importantly, existing process reward methods are often designed for isolated reasoning steps and lack explicit modeling of the logical dependencies. For example, a downstream brand-matching error may originate from an upstream step that incorrectly parses the query-side brand intent. Ignoring such dependency-driven error propagation limits the optimization of structured relevance reasoning.

To address fine-grained credit assignment in relevance reasoning, we propose Graph-GRPO, a reinforcement learning method based on graph-structured credit assignment. Graph-GRPO models the relevance judgment process as a reasoning dependency graph, where key components of query-product matching reasoning are represented as graph nodes and their logical dependencies are represented as graph edges. Within the GRPO \cite{shao2024deepseekmath,guo2025deepseek} framework, outcome-level rewards are propagated along the graph structure and converted into structured node-level feedback for different reasoning nodes. This enables fine-grained credit assignment from the global decision to local evidence. To improve graph-based credit assignment, we parameterize the propagation coefficients on graph edges as learnable variables and incorporate them into the computation of the main policy loss. These coefficients are therefore updated by gradients from the main loss, allowing the strength of credit propagation to align with the policy optimization objective. 

Around Graph-GRPO, we further adopt CoT random masking for supervised policy initialization and perform graph-node-based multi-head joint distillation to transfer structured reasoning ability and graph dependencies to a lightweight online model. Together, these modules form a trainable and deployable framework for generative relevance modeling. Extensive offline evaluations and online A/B tests demonstrate that the proposed framework delivers consistent improvements in both relevance quality and business-critical metrics, thereby enhancing the overall user experience of e-commerce search systems.

Our key contributions are summarized as follows:

\begin{itemize}

\item We propose Graph-GRPO, a graph-structured RL method for e-commerce search relevance modeling. It represents relevance reasoning as a dependency graph and propagates outcome-level rewards into node-level credit signals, enabling dependency-aware fine-grained optimization.

\item We introduce a main-loss-driven learnable graph coefficient mechanism that jointly updates edge propagation coefficients with the policy, allowing adaptive credit allocation across different reasoning paths.

\item We build a generative relevance modeling framework centered on Graph-GRPO, combining CoT random masking for stable initialization and graph-node-based multi-head distillation for lightweight deployment. Offline evaluations and online A/B tests show consistent gains in relevance classification, key business metrics, and user-facing search quality.

\end{itemize}

\section{Related Work}

\begin{figure*}[t]
\centering
\includegraphics[scale=0.51]{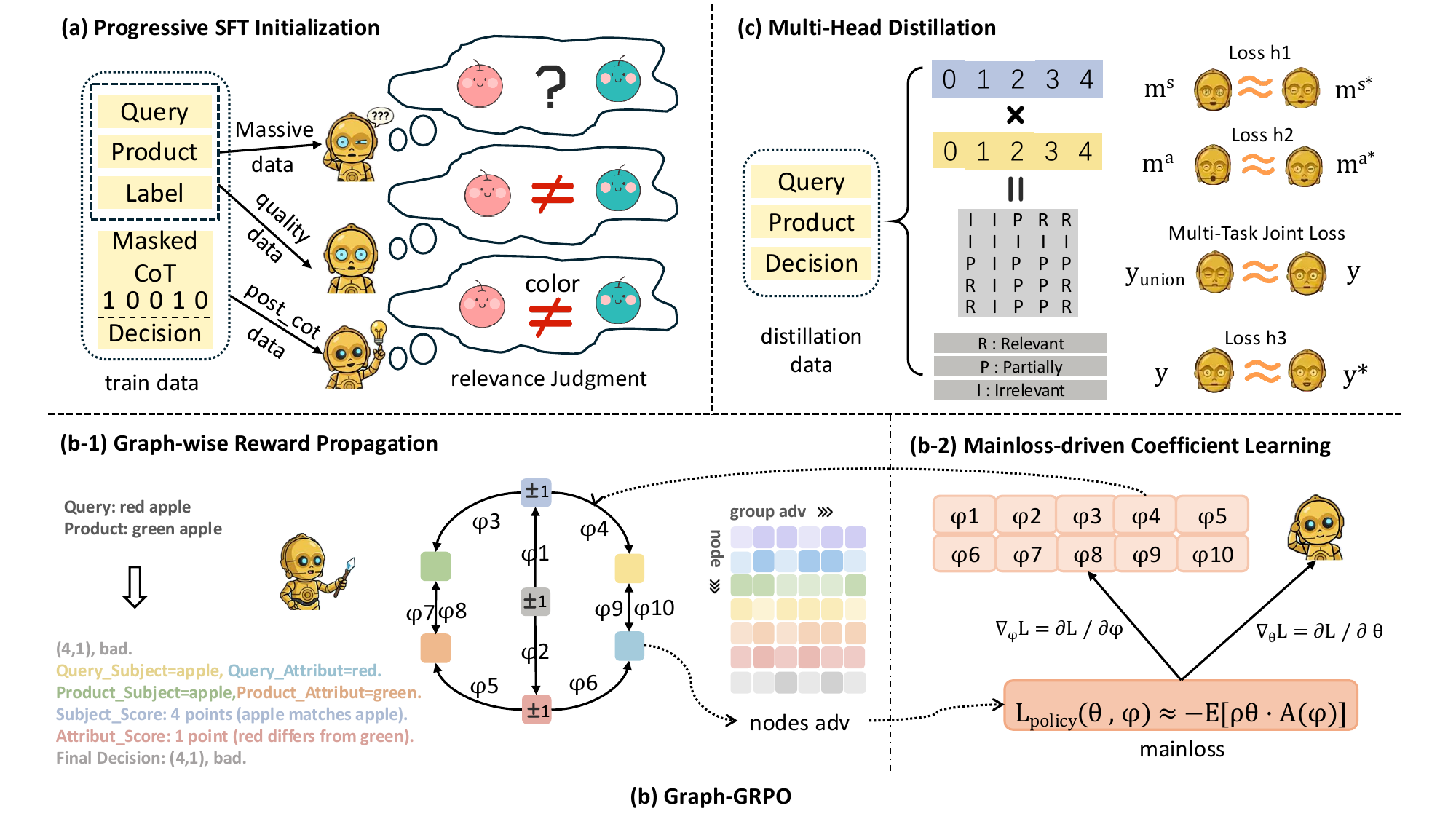}
\caption{Graph-GRPO-Centered Training and Deployment Framework for Generative Relevance Modeling. The framework consists of progressive SFT initialization for stable policy initialization, Graph-GRPO for dependency-aware reward propagation and main-loss-driven coefficient learning, and graph-node-based multi-head distillation for lightweight online deployment.}
\label{fig:Graph_grpo}
\end{figure*}

Relevance modeling is a core task in e-commerce search, aiming to assess the matching degree between a user query and a candidate product. Early methods mainly rely on lexical matching and handcrafted features, such as BM25 \cite{robertson1994okapi} and TF-IDF \cite{sparck2004statistical,salton1965smart}. These methods capture lexical overlap between queries and products, but are limited in modeling deeper semantic relevance. With deep learning \cite{lecun2015deep,hinton2006reducing,bengio2003neural}, DSSM \cite{huang2013learning} and related neural matching models \cite{shen2014latent,hu2014convolutional,pang2016text,xiong2017end} learn dense query-product representations, reducing the reliance on handcrafted features. Later, attention mechanisms \cite{bahdanau2014neural} and Transformer-based pre-trained encoders \cite{vaswani2017attention}, such as BERT \cite{devlin2019bert} and its variants \cite{liu2019roberta,lan2019albert,clark2020electra}, further improve contextual semantic modeling and become mainstream methods for relevance modeling.

Recently, autoregressive LLMs \cite{pradeep2023rankvicuna,zhuang2023rankt5,sun2023chatgpt} have advanced generative relevance modeling by leveraging strong language understanding and reasoning capabilities. LREF \cite{tang2025lref} adapts LLMs to product search through Multi-CoT fine-tuning and DPO-based debiasing, showing the potential of LLM-based relevance reasoning in industrial search scenarios. However, supervised fine-tuning and preference optimization still largely depend on fixed training data or preference pairs, which limits their ability to directly optimize generation policies using task-specific feedback.

This limitation further motivates reinforcement learning as a more suitable post-training paradigm for optimizing generation policies with task-specific feedback \cite{williams1992simple,schulman2015trust,schulman2017proximal,ouyang2022training,wei2021finetuned}. GRPO \cite{shao2024deepseekmath,guo2025deepseek} is a representative RL method that samples multiple responses for each input and estimates relative advantages through within-group reward normalization, thereby avoiding the need for an additional value model. TaoSR1 \cite{dong2025taosr1} improves generative relevance reasoning through a multi-stage training framework that combines pass@N-based preference optimization with difficulty-driven GRPO training.
However, existing GRPO-style relevance methods are still primarily driven by final-outcome feedback, where the generated reasoning process is optimized as a flat sequence. Such sequence-level credit assignment cannot distinguish faulty reasoning steps from correct intermediate ones, which can lead to misaligned credit assignment.

To mitigate coarse outcome-level reward propagation, SAM \cite{zeng2026optimizing} selectively reinforces or penalizes intermediate reasoning segments for more precise reward assignment. Other methods provide denser supervision through manual process rewards \cite{mu2024rule,xia2026reasoning} or external process reward models \cite{uesato2022solving,lightman2024let,wang2024math,luo2023wizardmath}. However, these process rewards are costly to construct and often unstable in practice. More importantly, existing methods usually handle reasoning steps independently and ignore dependency-driven error propagation, limiting the optimization of structured relevance reasoning.

Unlike these methods, Graph-GRPO models e-commerce search relevance judgment as a reasoning dependency graph and converts sparse outcome-level rewards into fine-grained node-level feedback through dependency-aware credit propagation. It further introduces a main-loss-driven learnable graph coefficient mechanism to learn task-aligned responsibility allocation and improve the effectiveness of credit propagation across reasoning paths.

\section{Preliminaries}

We formulate e-commerce search relevance modeling as a supervised multi-class classification task, with the objective of determining the relevance between a user query and a candidate product. Given a user query $Q$ and a candidate product $P$, the model learns a relevance decision function $f(Q,P) \rightarrow y$, where $y \in Y$. In the JD e-commerce search scenario, we adopt a three-level relevance annotation scheme, denoted as $Y=\{\mathrm{relevant}, \mathrm{partially\ relevant}, \mathrm{irrelevant}\}$. This label scheme reflects the extent to which a product satisfies the user's search intent and serves as the basis for subsequent generative relevance modeling and reinforcement learning optimization.
\section{Methodology}


This section presents a Graph-GRPO-centered framework for training and deploying generative relevance models. As shown in Fig.~\ref{fig:Graph_grpo}, the pipeline includes supervised policy initialization, graph-structured RL optimization, and lightweight model distillation for deployment.

First, we initialize a strong generative relevance model through two-stage progressive SFT with label-only supervision and CoT random masking. Then, Graph-GRPO parses model outputs into a reasoning dependency graph of reasoning steps and logical dependencies. Graph-structured reward propagation assigns fine-grained credit from the global decision to local evidence. Meanwhile, graph-edge propagation coefficients are learnable parameters jointly optimized by the main policy loss, enabling adaptive responsibility allocation across reasoning paths. Finally, we transfer the reasoning ability of the Graph-GRPO-optimized large model to a lightweight online model through graph-node-based multi-head distillation, meeting the low-latency requirements of industrial search systems.

\subsection{Progressive SFT Initialization}


Full CoT supervision during initialization may dilute relevance discrimination with weakly related explanatory tokens, causing the model to fit templated or noisy rationales rather than a stable decision boundary. Therefore, we adopt an easy-to-hard label-only SFT strategy, progressively training on large-scale model-annotated data and then high-quality human-annotated data, so the model learns stable final-decision ability before structured CoT reasoning.

After obtaining a stable label-only policy, we introduce structured reasoning supervision through masked CoT training. Since the previous SFT stage trains the model to output the relevance label directly, switching to a conventional reason-then-answer CoT format would change the generation order and place label supervision after a long reasoning segment. This may weaken the stable decision behavior learned from label-only SFT. Therefore, we adopt a label-first CoT format with an intuition-reasoning-decision order. The model first outputs a preliminary relevance label, then generates structured reasoning, and finally produces the final relevance decision. The preliminary label aligns the generation format with label-only SFT, while the final decision is used for subsequent relevance judgment and graph construction.

To prevent the long reasoning segment from dominating the SFT objective and weakening the supervision of the final decision, we apply stochastic loss masking only to reasoning tokens. The preliminary label and final decision are always supervised, while only a random subset of reasoning tokens contributes to the loss. This preserves stable decision learning, retains partial structured-reasoning supervision, and reduces overfitting to templated explanations.

Let the complete CoT output be
\(
o_i=[\tilde{y}_i\,;\,r_i\,;\,y_i],
\)
where \(\tilde{y}_i\) denotes the preliminary relevance label, \(r_i=(r_{i,1},\ldots,r_{i,T_i})\) denotes the reasoning content, and \(y_i\) denotes the final relevance decision. For each reasoning token \(r_{i,t}\), we sample a binary mask \(m_{i,t}\sim\mathrm{Bernoulli}(1-\rho)\), where \(\rho\) is the masking probability. The training objective is defined as

\begin{equation}
\begin{aligned}
\mathcal{L}_{\mathrm{CoT}\text{-}\mathrm{Masked}}
=&
-\sum_{i=1}^{N}
\left[
\log P_{\phi}(\tilde{y}_i\mid q_i,p_i)
+
\log P_{\phi}(y_i\mid q_i,p_i,\tilde{y}_i,r_i)
\right] \\
&-
\sum_{i=1}^{N}
\sum_{t=1}^{T_i}
m_{i,t}
\log P_{\phi}(r_{i,t}\mid q_i,p_i,\tilde{y}_i,r_{i,<t}) .
\end{aligned}
\end{equation}

\subsection{Graph-GRPO}
\subsubsection{Structured Relevance Reasoning}In e-commerce search relevance modeling, relevance judgment is not merely a lexical matching problem. It requires assessing whether the product subject matches the user's core purchase intent and whether attribute constraints such as brand, model, and size are satisfied.

Following this task logic, we formalize structured relevance reasoning into two dimensions: subject-level matching and attribute-level matching. Given a query-product pair \((q,p)\), the generative relevance model first parses the query and the product into four semantic components:
\(
(q,p)\rightarrow(q^{s},q^{a},p^{s},p^{a}),
\)
where \(q^{s}\) and \(p^{s}\) denote the query-side and product-side subject descriptions, respectively, and \(q^{a}\) and \(p^{a}\) denote the corresponding attribute descriptions. Based on these components, the model further predicts a subject-level matching score and an attribute-level matching score:
\begin{equation}
(m^{s},m^{a})
=
\left(
g^{s}(q^{s},p^{s}),
g^{a}(q^{a},p^{a})
\right),
\quad
m^{s},m^{a}\in\{0,1,2,3,4\}.
\end{equation} The five scores correspond to no intent, mismatch, weak match, partial match, and exact match, respectively. The final relevance label is determined by a mapping function over the two scores:
\(
y=h(m^{s},m^{a}), y\in Y.
\)
Thus, the complete relevance reasoning process can be summarized as:
\(
(q,p)\rightarrow
(q^{s},q^{a},p^{s},p^{a})
\rightarrow
(m^{s},m^{a})
\rightarrow y.
\)

This structured output naturally forms seven semantic units: the final label \({y}\), the subject-level score \(m^{s}\), the attribute-level score \(m^{a}\), and four evidence components \(q^{s}\), \(q^{a}\), \(p^{s}\), and \(p^{a}\). Graph-GRPO uses these semantic units as graph nodes, enabling RL to assign credit not only to the final decision, but also to the intermediate matching scores and the local evidence on which they depend.

\subsubsection{Dependency-aware Graph Credit Assignment}

Based on the structured reasoning process, Graph-GRPO constructs a reasoning dependency graph \(\mathcal{G}=(\mathcal{N},\mathcal{E})\) over seven semantic units:
\(
\mathcal{N}
=
\{y,m^{s},m^{a},q^{s},q^{a},p^{s},p^{a}\}.
\)
The graph edges explicitly encode the logical dependencies in relevance reasoning. The final decision depends on the subject-level and attribute-level matching scores, and each score is further grounded in the corresponding Query/Product-side evidence. We define the edge set as
\begin{equation}
\begin{array}{@{}l@{\;}c@{\;}l@{}}
\mathcal{E}
&=&
\mathcal{E}_{\mathrm{dec}}
\cup
\mathcal{E}_{\mathrm{evid}}
\cup
\mathcal{E}_{\mathrm{cons}},
\\[2pt]
\mathcal{E}_{\mathrm{dec}}
&=&
\{y\rightarrow m^{s},\; y\rightarrow m^{a}\},
\\[2pt]
\mathcal{E}_{\mathrm{evid}}
&=&
\{m^{s}\rightarrow q^{s},\; m^{s}\rightarrow p^{s},\;
m^{a}\rightarrow q^{a},\; m^{a}\rightarrow p^{a}\},
\\[2pt]
\mathcal{E}_{\mathrm{cons}}
&=&
\{q^{s}\leftrightarrow q^{a},\; p^{s}\leftrightarrow p^{a}\}.
\end{array}
\end{equation}
Here, \(\mathcal{E}_{\mathrm{dec}}\) propagates global decision feedback to dimension-level scores, \(\mathcal{E}_{\mathrm{evid}}\) propagates score-level responsibility to Query/Product-side evidence, and \(\mathcal{E}_{\mathrm{cons}}\) captures intra-side consistency between subject and attribute evidence on the Query side and Product side.

We assign exact-match base rewards to the final decision and the two matching scores. Specifically, \(r_y\), \(r_s\), and \(r_a\) denote the correctness rewards for the final decision, subject-level score, and attribute-level score, respectively. Let \(y^\ast\), \(m^{s,\ast}\), and \(m^{a,\ast}\) denote the corresponding gold annotations. The base rewards are defined as
\begin{equation}
r_y=2\mathbb{I}[y=y^\ast]-1,\quad
r_s=2\mathbb{I}[m^s=m^{s,\ast}]-1,\quad
r_a=2\mathbb{I}[m^a=m^{a,\ast}]-1,
\end{equation}
where \(\mathbb{I}[\cdot]\) equals 1 if the condition holds and 0 otherwise. Thus, each component receives \(+1\) for an exact match and \(-1\) otherwise.

To allow different dependency paths to carry different responsibility strengths, we assign an edge-wise propagation coefficient to each directed edge in the graph. Let
\(
\boldsymbol{\varphi}
=
\{\varphi_{1},\varphi_{2},\ldots,\varphi_{10}\}
\)
denote the propagation coefficients for the ten directed propagation edges. These coefficients control how much reward or penalty is transferred from an upstream node to a downstream node during graph-based credit propagation.

Here, \(r\) denotes the local exact-match reward of directly supervised nodes, \(R\) denotes the graph-propagated node reward, and \(\widehat{R}\) denotes the intermediate evidence reward before cross-dimension consistency aggregation.
The global decision reward is first propagated to the two matching-score nodes:
\begin{equation}
\begin{aligned}
R_y=r_y,\quad
R_{m^{s}}=r_s+\varphi_1 r_y,\quad
R_{m^{a}}=r_a+\varphi_2 r_y.
\end{aligned}
\end{equation}

The score-node rewards are then propagated to the corresponding evidence nodes as intermediate rewards:
\begin{equation}
\begin{aligned}
\widehat{R}_{q^{s}}=\varphi_3R_{m^{s}},
\quad
\widehat{R}_{p^{s}}=\varphi_4R_{m^{s}},
\quad
\widehat{R}_{q^{a}}=\varphi_5R_{m^{a}},
\quad
\widehat{R}_{p^{a}}=\varphi_6R_{m^{a}}.
\end{aligned}
\end{equation}

The intermediate evidence rewards then exchange subject-attribute consistency signals within the Query side and Product side:
\begin{equation}
\begin{aligned}
R_{q^{s}}&=\widehat{R}_{q^{s}}+\varphi_7\widehat{R}_{q^{a}},&
R_{q^{a}}&=\widehat{R}_{q^{a}}+\varphi_8\widehat{R}_{q^{s}},\\
R_{p^{s}}&=\widehat{R}_{p^{s}}+\varphi_9\widehat{R}_{p^{a}},&
R_{p^{a}}&=\widehat{R}_{p^{a}}+\varphi_{10}\widehat{R}_{p^{s}}.
\end{aligned}
\end{equation}
In this way, Graph-GRPO transforms sparse correctness signals from the final label and dimension-level matching scores into more targeted fine-grained node-level rewards through dependency-aware propagation over the reasoning graph.

\subsubsection{Node-wise GRPO}

After graph-based reward propagation, each sampled response receives node-level rewards. Given an input \(x_i=(q_i,p_i)\), the policy samples \(M\) responses \(\{o_{i,j}\}_{j=1}^{M}\). For each response \(o_{i,j}\) and node \(n\in\mathcal{N}\), let \(R_{i,j,n}\) denote the propagated reward of node \(n\). Instead of computing a single sequence-level advantage, Graph-GRPO normalizes rewards within each node:
\begin{equation}
\begin{aligned}
A_{i,j,n}
=
\frac{
R_{i,j,n}-\mu_{i,n}
}{
\sigma_{i,n}+\epsilon_{\mathrm{std}}
}
\end{aligned}
\end{equation}
where \(\mu_{i,n}\) and \(\sigma_{i,n}\) denote the mean and standard deviation of node \(n\)'s rewards among the \(M\) sampled responses for input \(x_i\).

To apply node-wise advantages to token-level policy optimization, we assign each token to one of the seven nodes according to the output template. Let
\(
(n_1,\ldots,n_7)=(y,m^{s},m^{a},q^{s},q^{a},p^{s},p^{a})
\)
denote the ordered node list, and let \(a_{i,j,t}\in\{0,1,\ldots,7\}\) denote the node index assigned to the \(t\)-th token in response \(o_{i,j}\), where \(a_{i,j,t}=0\) indicates non-semantic template tokens. The token-level advantage is defined as
\begin{equation}
A_{i,j,t}
=
\begin{cases}
A_{i,j,n_{a_{i,j,t}}}, & a_{i,j,t}>0,\\
0, & a_{i,j,t}=0.
\end{cases}
\end{equation}
The probability ratio and its clipped version are defined as
\begin{equation}
\begin{aligned}
\rho_{i,j,t}
=
\frac{
\pi_{\theta}
\left(
o_{i,j,t}\mid o_{i,j,<t},x_i
\right)
}{
\pi_{\theta_{\mathrm{old}}}
\left(
o_{i,j,t}\mid o_{i,j,<t},x_i
\right)
},
\quad
\bar{\rho}_{i,j,t}
=
\operatorname{clip}
\left(
\rho_{i,j,t},
1-\epsilon_{\mathrm{clip}},
1+\epsilon_{\mathrm{clip}}
\right).
\end{aligned}
\end{equation}
We add KL regularization to limit drift from the reference model:
\begin{equation}
\begin{aligned}
\mathcal{K}_{i,j,t}
=
D_{\mathrm{KL}}
\left(
\pi_{\theta}(\cdot\mid o_{i,j,<t},x_i)
\,\|\, 
\pi_{\mathrm{ref}}(\cdot\mid o_{i,j,<t},x_i)
\right).
\end{aligned}
\end{equation}
The KL-regularized Graph-GRPO objective is:
\begin{equation}
\begin{aligned}
\mathcal{L}_{\mathrm{Graph}\text{-}\mathrm{GRPO}}
=
-\mathbb{E}_{i,j,t}
\left[
\min
\left(
\rho_{i,j,t}A_{i,j,t},
\bar{\rho}_{i,j,t}A_{i,j,t}
\right)
-
\beta \mathcal{K}_{i,j,t}
\right],
\end{aligned}
\end{equation}
This objective preserves the group-relative comparison mechanism of GRPO while replacing sequence-level credit assignment with node-level credit assignment. Consequently, the policy update is guided by node-specific advantages, allowing Graph-GRPO to optimize different reasoning units separately rather than applying a uniform advantage to the entire response.

\subsubsection{Main-Loss-Driven Graph Coefficients}

In graph-structured credit assignment, rewards and penalties should not be propagated equally along all edges. Different reasoning paths may carry different levels of responsibility for the final prediction. For example, an incorrect final decision may be mainly caused by an error in subject-level matching, or by an attribute mismatch. Therefore, using fixed propagation coefficients is insufficient to adapt to different training stages, sample distributions, and error types.

To address this issue, we introduce a main-loss-driven learnable graph coefficient mechanism. We parameterize the propagation coefficient of each graph edge as a learnable variable:
\(
\boldsymbol{\varphi}
=
\{\varphi_{e}\mid e\in\mathcal{E}\}.
\)
To keep responsibility allocation stable and interpretable, coefficients are normalized within each branch group. For an edge \(e=(u\rightarrow v)\), its normalized coefficient is defined as
\begin{equation}
\begin{aligned}
w_{u\rightarrow v}
=
\frac{
\exp(\varphi_{u\rightarrow v})
}{
\sum_{e'\in\mathcal{B}(e)}
\exp(\varphi_{e'})
},
\end{aligned}
\end{equation}
where \(\mathcal{B}(e)\) denotes the branch group of edge \(e\), consisting of edges that compete to distribute the same upstream credit, such as \(y\rightarrow m^{s}\) and \(y\rightarrow m^{a}\). This normalization makes credit allocation within each group bounded and comparable, preventing unstable growth of the reward scale.

With learnable coefficients, reward propagation over the graph is written as a weighted message-passing process:
\begin{equation}
\begin{aligned}
R_v
=
r_v
+
\sum_{u\in\mathcal{P}(v)}
w_{u\rightarrow v}R_u,
\end{aligned}
\end{equation}
where \(r_v\) is the base reward of node \(v\), \(\mathcal{P}(v)\) is the set of upstream nodes, and \(R_u\) is the propagated reward from node \(u\). Thus, the final node reward is controlled by learnable edge-wise propagation coefficients.

Unlike training a separate reward-side learner, we directly optimize the graph coefficients through the main policy loss. Let \(\theta\) denote the policy parameters and \(\boldsymbol{\varphi}\) denote the learnable graph coefficients. The training objective is
\begin{equation}
\begin{aligned}
\mathcal{J}(\theta,\boldsymbol{\varphi})
=
\mathcal{L}_{\mathrm{Graph}\text{-}\mathrm{GRPO}}(\theta,\boldsymbol{\varphi})
+
\lambda_{\mathrm{reg}}\mathcal{R}(\boldsymbol{\varphi}).
\end{aligned}
\end{equation}
where \(\lambda_{\mathrm{reg}}\) controls the regularization strength, and \(\mathcal{R}(\boldsymbol{\varphi})\) regularizes the learnable graph coefficients to stabilize coefficient learning and credit propagation.
We jointly optimize \(\theta\) and \(\boldsymbol{\varphi}\) under this objective. Since node rewards, node-wise advantages, and token-level policy losses depend on the propagation coefficients, gradients from the main policy loss can update \(\boldsymbol{\varphi}\) together with the policy.

This design integrates graph coefficients into the policy optimization process. Because the coefficients affect how node rewards are propagated to the policy loss, gradients from the main policy loss can strengthen effective credit propagation paths while suppressing noisy ones. Consequently, Graph-GRPO learns edge-wise responsibility allocation according to the policy improvement objective, rather than relying on fixed propagation rules.

\subsection{Graph-node Multi-head Distillation}

The Graph-GRPO optimized LLM provides strong structured reasoning ability, but its inference cost makes it difficult to deploy directly in online search systems. To meet the latency requirement of industrial e-commerce search, we distill the teacher model into a lightweight cross-encoder BERT student.

Given a query-product pair $(q,p)$, the student model produces a shared representation $h_{\psi}$ and uses three prediction heads to model the key graph nodes:
\begin{equation}
\begin{aligned}
m^{s}=h_{1}(h_{\psi}),\qquad
m^{a}=h_{2}(h_{\psi}),\qquad
y=h_{3}(h_{\psi}),
\end{aligned}
\end{equation}
where $m^{s}$ and $m^{a}$ are the subject-score and attribute-score predictions, while $y$ is the final relevance prediction. Let $m^{s\ast}$ and $m^{a\ast}$ denote the gold subject and attribute scores, and let $p_y^\ast$ denote the softened one-hot distribution of the gold relevance label. The losses are:
\begin{equation}
\begin{aligned}
\mathcal{L}_{s}=\mathrm{CE}(m^{s},m^{s\ast}),\quad
\mathcal{L}_{a}=\mathrm{CE}(m^{a},m^{a\ast}),\quad
\mathcal{L}_{y}=\mathrm{KL}(p_y^\ast \parallel P(y)).
\end{aligned}
\end{equation}

Since the final relevance label is determined by the subject score and the attribute score under the annotation rule, independently training the three heads is insufficient to explicitly capture their internal reasoning relationship. Based on this task logic, we derive an induced relevance prediction from the two score heads:
\(
y_{\mathrm{union}}=g(m^{s},m^{a}),
\)
where $g(\cdot)$ maps the predicted subject and attribute scores to the final relevance label according to the annotation rule. We then introduce a consistency loss to align the induced relevance prediction with the direct prediction from the result head:
\begin{equation}
\begin{aligned}
\mathcal{L}_{\mathrm{c}}=
\mathrm{KL}
\left(
P(y_{\mathrm{union}})
\parallel
P(y)
\right).
\end{aligned}
\end{equation}

The overall training objective is:
\begin{equation}
\begin{aligned}
\mathcal{L}_{\mathrm{distill}}
=
\lambda_{s}\mathcal{L}_{\mathrm{s}}
+
\lambda_{a}\mathcal{L}_{\mathrm{a}}
+
\mathcal{L}_{\mathrm{y}}
+
\lambda_{c}\mathcal{L}_{\mathrm{c}}.
\end{aligned}
\end{equation}
where \(\lambda_{s}\), \(\lambda_{a}\), and \(\lambda_{c}\) are balancing coefficients for the subject, attribute, and consistency losses, respectively, with the result loss used as the main supervision term.

Through graph-node multi-head distillation, the student model learns not only the final relevance label and the two dimension-level scores, but also the logical consistency between dimension scores and the final decision. This method converts the structured reasoning signals from the LLM teacher into learnable multi-task supervision for the lightweight student model, enabling the online model to achieve stronger structured relevance modeling while maintaining efficient inference.

\begin{table}[t]
\centering
\caption{Distribution of the Test Dataset}
\label{tab:dataset_distribution}
\begin{tabular}{lcc}
\toprule
Relevance Level & Number & Proportion \\
\midrule
Irrelevant & 27,640 & 29.91\% \\
Partially Relevant & 20,418 & 22.09\% \\
Relevant & 44,361 & 48.00\% \\
\midrule
\textbf{Total} & \textbf{92,419} & \textbf{100.00\%} \\
\bottomrule
\end{tabular}
\end{table}

\begin{table*}[t]
\centering
\caption{Evaluation Results}
\label{tab:evaluation_results}
\resizebox{\textwidth}{!}{
\begin{tabular}{l|ccc|ccc}
\toprule
\multirow{2}{*}{\textbf{Models}} 
& \multicolumn{3}{c|}{\textbf{Three-Level Classification}} 
& \multicolumn{3}{c}{\textbf{Irrelevant Classification}} \\
& \textbf{Macro F1} & \textbf{Weighted F1} & \textbf{Accuracy}
& \textbf{Precision} & \textbf{Recall} & \textbf{F1} \\
\midrule
LLM-Base 
& 56.39 & 61.34 & 61.54 
& \textbf{84.58} & 29.70 & 43.97 \\

\midrule

LLM-SFT w/o Random Masking
& 79.55 & 82.47 & 83.19  
& 76.61 & 90.71 & 83.06 \\

LLM-SFT
& 80.33 & 83.13 & 83.84 
& 77.95 & \textbf{90.94} & 83.94 \\

\midrule

LLM-GRPO (based on LLM-SFT)
& 80.91 & 83.57 & 84.21 
& 79.76 & 89.90 & 84.53 \\

\midrule

Graph-GRPO w/o Learnable Coefficients
& 81.22 & 83.80 & 84.41
& 80.39 & 89.57 & 84.73 \\

Graph-GRPO
& \textbf{81.26} & \textbf{83.84} & \textbf{84.44}
& 80.11 & 89.95 & \textbf{84.75} \\
\bottomrule

\end{tabular}
}
\end{table*}
\section{Experiments}
\subsection{Datasets and Metrics}
\noindent\textbf{Datasets:}
The test set is drawn from real JD search logs, containing over 90,000 query-product pairs across diverse e-commerce search scenarios. Each instance is manually verified in multiple rounds and labeled as relevant, partially relevant, or irrelevant. Table~\ref{tab:dataset_distribution} reports the test-set label distribution.

\noindent\textbf{Metrics:}
We evaluate multi-class classification performance from two perspectives. First, for the three-level relevance classification task, we report Accuracy, Macro-F1, and Weighted-F1 to evaluate the model's overall ability to discriminate among different relevance levels. Strong three-level classification performance indicates that the model can more reliably distinguish different degrees of relevance. This is important for search ranking, where highly relevant products should be prioritized and boundary errors between partially relevant and irrelevant samples should be reduced. Second, since irrelevant product exposure directly affects user experience, we further report Precision, Recall, and F1-Score for the irrelevant class to evaluate the model's ability to identify and filter irrelevant products. These class-wise metrics provide an offline measure of irrelevant-case detection and are practically important for reducing the exposure of irrelevant products in top-ranked positions.

\subsection{Implementation Detail}
We build Graph-GRPO based on Qwen3-14B \cite{qwen2025qwen314b}. In the label-only SFT stage, the learning rate is \(1\times10^{-5}\), with weight decay \(0.1\). In the CoT masking SFT stage, the learning rate is \(7\times10^{-6}\), with weight decay \(0.1\) and maximum sequence length \(1024\). In the Graph-GRPO stage, the learning rate is \(1\times10^{-6}\), with weight decay \(0.1\), training batch size \(512\), and PPO mini-batch size \(256\). For each sample, the policy model generates \(16\) responses at a sampling temperature of \(0.95\) for within-group relative advantage estimation. For learnable graph coefficients, the target learning rate is \(6\times10^{-4}\), the regularization weight is \(6\times10^{-5}\), and gradient clipping with maximum norm \(6.0\) is applied to stabilize coefficient updates.
\subsection{Baselines}
We compare Graph-GRPO with several representative baselines:

\noindent \textbf{LLM-Base:} The vanilla Qwen3-14B model is directly evaluated to establish its zero-shot performance on the relevance task.

\noindent \textbf{LLM-SFT:} The model is trained with two-stage label-only SFT and CoT random masking SFT, serving as the CoT-SFT baseline.

\noindent \textbf{LLM-GRPO:} Based on LLM-SFT, standard GRPO optimizes the generation policy with outcome-level rewards as the conventional RL baseline.

\noindent \textbf{Graph-GRPO (Ours):} Our method extends LLM-SFT with dependency-aware graph reward propagation, node-wise advantage estimation, and learnable graph coefficients for fine-grained credit assignment.

\subsection{Offline Evaluation}
Table~\ref{tab:evaluation_results} reports offline evaluation results.
The vanilla LLM Base shows poor zero-shot performance on the relevance task, achieving only 56.39\% Macro-F1. More importantly, although it obtains relatively high Precision on the irrelevant class, its Recall and F1 are only 29.70\% and 43.97\%, respectively. This indicates that the model identifies only a small portion of obvious irrelevant cases and lacks reliable task-specific calibration.

After CoT random masking SFT, LLM-SFT significantly improves overall classification performance, reaching 80.33\% Macro-F1. This demonstrates the importance of task-specific supervised initialization for generative relevance modeling. However, its irrelevant-class performance remains imbalanced. It achieves a high Recall of 90.94\% on the irrelevant class, but its Precision is only 77.95\% and its F1 is 83.94\%, suggesting that LLM-SFT tends to over-predict irrelevant cases. Standard GRPO further optimizes the generation policy with outcome-level rewards on top of LLM-SFT and brings more balanced improvements. Nevertheless, as it treats the whole response as a single optimization unit, the performance gain remains limited.

Graph-GRPO achieves the best overall performance among all methods, with the highest Macro-F1 and irrelevant-class F1 of 81.26\% and 84.75\%, respectively. Compared with standard GRPO, Graph-GRPO yields consistent gains across all offline metrics, covering both overall three-level relevance classification and irrelevant-class detection. This demonstrates that the proposed graph-based optimization scheme provides a comprehensive enhancement over conventional outcome-level GRPO. The results further suggest that assigning rewards according to reasoning dependencies offers more informative supervision than treating the entire generated response as a single optimization unit, leading to more robust relevance prediction and more reliable irrelevant-product detection.

\begin{figure*}[t]
    \centering
    \includegraphics[width=0.85\textwidth]{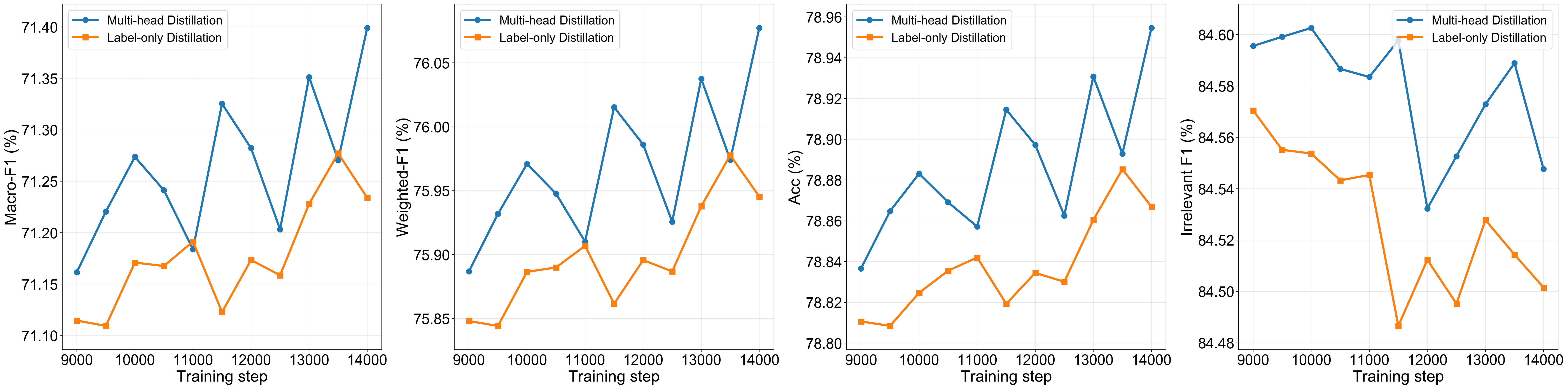}
    \caption{Ablation results of graph-node-based multi-head distillation on the lightweight online model.}
    \label{fig:distillation_ablation}
\end{figure*}

\subsection{Ablation Study}

Ablation studies assess key components of our framework.

\noindent\textbf{CoT random masking.}
As shown in Table~\ref{tab:evaluation_results}, CoT random masking outperforms the LLM-SFT variant without random masking across all reported metrics. It improves Macro-F1 from 79.55\% to 80.33\% and irrelevant-class F1 from 83.06\% to 83.94\%. This suggests that random masking prevents long or templated explanations from dominating SFT while preserving useful reasoning supervision. It therefore improves relevance discrimination and provides a better initial policy for reinforcement learning.

\noindent\textbf{Learnable graph coefficients.}
As shown in Table~\ref{tab:evaluation_results}, learnable graph coefficients improve all metrics except irrelevant-class Precision compared with the fixed-coefficient variant. This shows that main-loss-driven coefficient learning better adjusts credit propagation across reasoning paths and strengthens graph-structured credit assignment. Notably, irrelevant-class Recall increases from 89.57\% to 89.95\%, indicating that adaptive coefficients help the model capture key subject-level or attribute-level mismatch signals. This improves irrelevant-case detection and helps reduce irrelevant product exposure in search results.

\noindent\textbf{Graph-node-based multi-head distillation.}
As shown in Fig.~\ref{fig:distillation_ablation}, graph-node-based multi-head distillation consistently outperforms label-only distillation in Macro-F1, Weighted-F1, Accuracy, and irrelevant-class F1 throughout training. This suggests that final-label supervision alone cannot fully transfer the teacher model's structured relevance reasoning ability. By introducing additional supervision from subject-level and attribute-level graph nodes, multi-head distillation provides richer intermediate signals and enables the lightweight student model to better capture the matching logic behind the final decision. The gain in irrelevant-class F1 further shows its effectiveness in irrelevant-case detection, which is important for reducing irrelevant product exposure in online search.

\subsection{Online Evaluation}

To meet the latency and throughput requirements of production search systems, we use the Graph-GRPO optimized LLM as the teacher model and deploy a lightweight BERT student through graph-node-based multi-head distillation. We conduct an online A/B test in the JD search engine, where the treatment slot uses the Graph-GRPO distilled BERT model and two independent control slots, Base1 and Base2, use the production baseline.

We adopt a controlled query-level comparison protocol with 10,000 representative queries covering diverse e-commerce categories. The same query set is routed to all three slots, reducing the influence of traffic variation and query distribution shift. We use Bad Case Rate as the primary online metric. It is defined as the proportion of queries whose top-10 results contain at least one product labeled as irrelevant \cite{cao2026prectr,li2024multi,luo2020alicoco}. This metric directly reflects user-facing relevance quality because irrelevant products in top-ranked positions can substantially degrade the search experience.

Base1 and Base2 achieve similar Bad Case Rates of \(11.99\%\) and \(11.95\%\), confirming the stability of the control setting. The Graph-GRPO distilled model reduces the Bad Case Rate to \(11.65\%\), corresponding to a relative reduction of \(2.67\%\) over the average baseline rate of \(11.97\%\). These online results show that the improvements brought by Graph-GRPO can be effectively transferred to a lightweight deployable model and lead to meaningful quality gains in an industrial-scale search system.
\section{Conclusion}
This paper presents Graph-GRPO, a dependency-aware RL method for generative e-commerce search relevance modeling. It represents structured relevance reasoning as a dependency graph and propagates sparse outcome-level rewards into fine-grained node-level credit signals, enabling more precise optimization of reasoning components. It further introduces a main-loss-driven graph coefficient mechanism that adapts credit propagation across reasoning paths. With CoT random masking for stable policy initialization and graph-node-based multi-head distillation for lightweight deployment, Graph-GRPO forms a practical industrial relevance modeling framework. Extensive offline experiments and online A/B tests show improved relevance classification performance and user-facing search quality. Deployed in the JD e-commerce search system, the framework serves hundreds of millions of users and validates RL for industrial-scale LLM-based reasoning systems.
\section{GenAI Usage Disclosure}
Generative AI tools were used only for minor language editing, including grammar, clarity, and readability improvements. They were not used to generate scientific content, ideas, methods, results, analyses, interpretations, or conclusions. The authors reviewed all edits and take full responsibility.

\bibliographystyle{ACM-Reference-Format}
\bibliography{bibfile}

\appendix









\end{document}